\documentclass{amsart}

\usepackage{amsmath,amssymb,amsbsy,amsthm,amsfonts,mathtools,dsfont,hyperref,mathrsfs}

\usepackage{enumitem}
\usepackage{wrapfig}
\usepackage[top=3.5cm,bottom=3.5cm,left=3.5cm,right=3.5cm]{geometry}
\usepackage[normalem]{ulem}

\usepackage{graphicx}
\usepackage{tikz}
\usepackage{tikz}
\usepackage{tikzscale}
\usetikzlibrary{math}
\usepackage{tikzscale}
\usetikzlibrary{intersections,decorations.markings}
\tikzset{>=latex}
\usepackage{pgfplots}
\pgfplotsset{compat=newest}
\usepgfplotslibrary{fillbetween}
\usepackage[normalem]{ulem}

\allowdisplaybreaks[3]

\newcommand{\tr}{\operatorname{Tr}}

\begin{document}

\newcommand{\ddd}{\,{\rm d}}

\def\os#1{{\color{blue}#1}}
\def\note#1{\marginpar{\small #1}}
\def\tens#1{\pmb{\mathsf{#1}}}
\def\vec#1{\boldsymbol{#1}}
\def\norm#1{\left|\!\left| #1 \right|\!\right|}
\def\fnorm#1{|\!| #1 |\!|}
\def\abs#1{\left| #1 \right|}
\def\ti{\text{I}}
\def\tii{\text{I\!I}}
\def\tiii{\text{I\!I\!I}}

\newcommand{\loc}{{\rm loc}}
\def\diver{\mathop{\mathrm{div}}\nolimits}
\def\grad{\mathop{\mathrm{grad}}\nolimits}
\def\Div{\mathop{\mathrm{Div}}\nolimits}
\def\Grad{\mathop{\mathrm{Grad}}\nolimits}
\def\cof{\mathop{\mathrm{cof}}\nolimits}
\def\det{\mathop{\mathrm{det}}\nolimits}
\def\lin{\mathop{\mathrm{span}}\nolimits}
\def\pr{\noindent \textbf{Proof: }}

\def\pp#1#2{\frac{\partial #1}{\partial #2}}
\def\dd#1#2{\frac{\d #1}{\d #2}}
\def\vec#1{\boldsymbol{#1}}

\def\0{\vec{0}}
\def\A{\mathcal{A}}
\def\B{\mathcal{B}}
\def\b{\vec{b}}
\def\C{\mathcal{C}}
\def\c{\vec{c}}
\def\vv{\vec{v}}
\def\DD{\vec{D}}
\def\Dv{\vec{D}\vv}
\def\BB{\vec{B}}
\def\e{\varepsilon}
\def\er{\epsilon}
\def\f{\vec{f}}
\def\F{\vec{F}}
\def\tF{\tilde{\F}}
\def\g{\vec{g}}
\def\G{\vec{G}}
\def\cG{\mathcal{\G}}
\def\H{\vec{H}}
\def\cH{\mathcal{H}}
\def\I{\vec{I}}
\def\Im{\text{Im}}
\def\j{\vec{j}}
\def\J{\vec{J}}
\def\dd{\vec{d}}
\def\k{\vec{k}}
\def\n{\vec{n}}
\def\q{\vec{q}}
\def\S{\vec{S}}
\def\s{\vec{s}}
\def\T{\vec{T}}
\def\u{\vec{u}}
\def\vp{\vec{\varphi}}
\def\vvt{\vv_{\tau}}
\def\vov{\vv\otimes\vv}
\def\cV{\mathcal{V}}
\def\w{\vec{w}}
\def\W{\vec{W}}
\def\x{\vec{x}}
\def\z{\vec{Z}}
\def\tz{\tilde{\z}}
\def\Z{\vec{Z}}
\def\X{\vec{X}}
\def\Y{\vec{Y}}
\def\balfa{\vec{\alpha}}

\def\Ge{\G_{\e}}
\def\ge{\g_{\e}}
\def\fidv{\phi_{\delta}(|\vv|^2)}
\def\fidve{\phi_{\delta}(|\ve|^2)}
\def\fidvd{\phi_{\delta}(|\vd|^2)}

\def\Ae{\A_\e}
\def\Aee{\Ae^\e}
\def\Aeetilde{\tilde{A}_\e^\e}
\def\Be{\B_\e}
\def\Bee{\Be^\e}
\def\De{\DD\ve}
\def\DDe{\DD^\e}
\def\Dvd{\DD\vd}
\def\oD{\overline{\DD}}
\def\tD{\tilde{\DD}}
\def\Dn{\DD^\e}
\def\Dno{\overline{\Dn}}
\def\Dnt{\tilde{\Dn}}
\def\Dm{\DD^\eta}
\def\Dmo{\overline{\Dm}}
\def\Dmt{\tilde{\Dm}}
\def\Se{\S^\e}
\def\se{\s^\e}
\def\ose{\overline{\se}}
\def\oS{\overline{\S}}
\def\tS{\tilde{\S}}
\def\Sn{\S^\e}
\def\Sno{\overline{\Sn}}
\def\Snt{\tilde{\Sn}}
\def\Sm{\S^\eta}
\def\Smo{\overline{\Sm}}
\def\Smt{\tilde{\Sm}}
\def\ve{\vv^\e}
\def\ove{\overline{\ve}}
\def\vove{\ve\otimes\ve}
\def\vd{\vv^\delta}
\def\sd{\s^\delta}
\def\Sd{\S^\delta}
\def\Dd{\DD\vd}

\def\Wnd#1{W^{1,#1}_{\n, \diver}}
\def\Wndr{W^{1,r}_{\n, \diver}}

\def\o{\Omega}
\def\po{\partial \Omega}
\def\dt{\frac{\d}{\d t}}
\def\pt{\partial_t}
\def\it{\int_0^t \!}
\def\iT{\int_0^T \!}
\def\io{\int_{\o} \!}
\def\iq{\int_{Q} \!}
\def\iqt{\int_{Q^t} \!}
\def\ipo{\int_{\po} \!}
\def\ig{\int_{\Gamma} \!}
\def\igt{\int_{\Gamma^t} \!}

\def\d{\, \textrm{d}}

\def\mn{\mathcal{P}}
\def\du{\mathcal{W}}
\def\tr{\operatorname{tr}}
\def\tow{\rightharpoonup}


\newtheorem{theorem}{Theorem}[section]
\newtheorem{lemma}[theorem]{Lemma}
\newtheorem{proposition}[theorem]{Proposition}
\newtheorem{remark}[theorem]{Remark}
\newtheorem{corollary}[theorem]{Corollary}
\newtheorem{definition}[theorem]{Definition}
\newtheorem{example}[theorem]{Example}

\numberwithin{equation}{section}

\title[On determination of boundary conditions]{On a methodology to determine whether the fluid slips adjacent to a solid surface}

\thanks{J.~M\'{a}lek acknowledges the support of the project No. 23-05207S financed by the Czech Science
Foundation (GA \v{C}R). J. M\'{a}lek is a member of the Ne\v{c}as Center for Mathematical Modelling. K~R.~Rajagopal thanks the Office of Naval Research for its support of this work.}

\author[J.~M\'{a}lek]{J.~M\'{a}lek}
\address{Charles University, Faculty of Mathematics and Physics, Mathematical institute, Sokolovsk\'{a} 83, 18675 Prague 8, Czech Republic}
\email{malek@karlin.mff.cuni.cz}

\author[K.~R.~Rajagopal]{K.~R.~Rajagopal}
\address{Department of Mechanical Engineering,  
Texas A\&M University, College Station, TX 77845 USA}
\email{krajagopal@tamu.edu}


\date{\today}

\begin{abstract} 
We discuss a methodology that could be gainfully exploited using easily measurable experimental quantities to ascertain if the ``no-slip" boundary condition is appropriate for the flows of fluids past a solid boundary.
\end{abstract}

\maketitle

\bigskip

\section{Introduction} \label{Sec1}

The imprimatur of Stokes popularized the assumption of the ``no-slip"\footnote{The notion of ``no-slip" at a solid boundary can be traced back to Daniel Bernoulli in Hydrodynamica.} boundary condition for the flow of a Navier-Stokes fluid adjacent to a solid boundary at the point of contact, even though Stokes was far from sanguine about its aptness, for Stokes \cite{Stokes1845} remarks: \emph{``Du Buat found by experiments that when the mean velocity of water flowing through a pipe is less than one inch in a second, the water near the inner surface of the pipe is at rest."} The speed at which the fluid is flowing, which Stokes is referring to, is very slow, and Stokes' opinion was that probably in such exceedingly slow flows the ``no-slip" boundary condition for the fluid adjacent to a solid
boundary might be appropriate.  In fact, Stokes also remarks: \emph{``The most interesting questions connected with this subject require for their solution a knowledge of the conditions which must be satisfied at the surface of a solid in contact with the fluid, which, except in the case of very small motions, are unknown."} The key part of the above quote from Stokes \emph{``\dots except in the case of very small motions, are unknown."} makes it abundantly clear that Stokes was far from convinced that the ``no-slip"  boundary condition for a fluid flowing past a solid boundary at the point of contact was felicitous in general flows. 

To determine experimentally whether the fluid slips, or does not slip at the boundary is challenging. While one could possibly make measurements very close to the boundary, to get data immediately adjacent to the boundary
is exceedingly difficult. There are numerous experiments, see for example \cite{Churaevetal1984, Vinogradova1999, Pitet2000, Craigetal2001, Baundryetal2001} amongst many others, that determine the slip for the fluids at the boundary. Moreover, the ``no-slip" boundary condition does not seem to be applicable to gases. It is also well known that the ``no-slip" condition is not applicable for the flows of several polymers adjacent to solid boundaries, see \cite{H1, H2}.

While our discussion in the paper is mainly concerned with the flows of Navier-Stokes fluids, in order to understand the main thesis of the work, it is best to set our discussion within the  context of a larger class of fluids. 

The main justification for the ``no-slip" assumption is supposedly the accordance of experimental predictions for a wide range of problems. That being said, the ``no-slip" condition is an \textbf{assumption}, no more and no less, and it is not
based on physical considerations. In fact, the early
savants, such as Coulomb, Girard, Navier, Poisson, Prony and others had competing hypotheses based on physical considerations, see Goldstein \cite{goldstein1938}. 

In view of the ``no-slip" condition merely being an assumption, three questions immediately offer themselves: 
\begin{itemize}
    \item Can we obtain solutions for flow problems which involve solid boundaries wherein we refrain from making the ``no-slip" assumption?
    \item Based on the above solution, can we assess the validity of the ``no-slip" boundary condition?
    \item What could have prompted the eager acceptance of the ``no-slip" condition by both fluid dynamicists and mathematicians? 
\end{itemize} 
In this short paper, we address all these questions. Let us address the last question first. 

\section{Preliminary considerations} 

A fluid is referred to as Stokesian fluid if the Cauchy stress $\T$ is a function of the density $\varrho$ and the symmetric part of the velocity gradient $\DD$ given by 
\begin{equation}
    \T = \vec{f}(\varrho, \DD)\,, \label{Jo}
\end{equation}
where 
\begin{equation}
\DD:=\frac12\left[\nabla \vec{v} + (\nabla \vec{v})^{T})\right],  \quad \vec{v} \textrm{ being the velocity.} \label{J1.3}
\end{equation}

We notice that the Navier-Stokes fluids and power-law fluids are subclasses of Stokesian fluids. In the case of incompressible Stokesian fluids, the constitutive relation takes the form 
\begin{equation}
    \T = -p\I + \vec{g}(\DD)\,. \label{Joo}
\end{equation}
We notice that \eqref{Jo} is a special subclass of implicit constitutive relations of the form (see Rajagopal \cite{Raj.2003,Raj.2006}) 
\begin{equation}
    \tilde{\!\!\vec{f}} (\varrho, \T, \DD) = 0\,, \label{Js}
\end{equation}
the incompressible counterpart being 
\begin{equation}
    \T = -p\I + \tilde{\vec{g}}(\T, \DD)\,. \label{Jss}
\end{equation}
Classification of incompressible fluids described by the implicit constitutive equations can be found in \cite{blechta2020}.

The Navier-Stokes fluids are characterized by the linear relation between the Cauchy stress $\T$ and $\DD$. 
More precisely, the constitutive equation for the compressible and the incompressible Navier-Stokes fluid are expressed, respectively, through
\begin{align}
\T &= -p(\varrho)\I + \lambda(\varrho)(\operatorname{tr} \DD) \I + 2\mu (\varrho) \DD,    \label{J1.2a}
\end{align}
and 
\begin{align}
\T &= -p\I + 2\mu \DD.    \label{J1.2b}
\end{align}
where in the former case $\varrho$ is the density, $p(\varrho)$ is the thermodynamic pressure, $\mu(\varrho)$ is the shear viscosity and $\lambda(\varrho)$ is referred to as the second coefficient of viscosity ($3\lambda + 2\mu$ is referred to as bulk modulus/bulk viscosity), while in the latter case $p\I$ is the  (constitutively) indeterminate part of the stress due to the constraint of incompressibility and $\mu$ is the constant shear viscosity.   

For the sake of illustrating the answer to the third question above, let us consider \eqref{J1.2b}. On substituting \eqref{J1.2b} into the balance of linear momentum
\begin{align}
    \varrho \frac{\textrm{d}\vec{v}}{\textrm{d}t} &= \diver \T + \varrho \vec{b}\,, \label{J1.4}
\end{align}
where $\vec{b}$ is the specific body force, and taking into cognizance that the fluid is incompressible, we obtain 
\begin{align}
    \diver \vec{v} & = 0\,, \label{J1.5} \\
    \varrho \frac{\textrm{d}\vec{v}}{\textrm{d}t} &= - \nabla p + \mu \Delta \vec{v} + \varrho \vec{b}\,, \label{J1.6}
\end{align}
for the two unknowns $\vec{v}$ and $p$. It would thus seem natural to prescribe the boundary conditions for the velocity $\vec{v}$. Taking the $\operatorname{curl}$ of equation \eqref{J1.6}, one obtains, in the case of a conservative body forces, a partial differential equation of higher order in the velocity, but the pressure is eliminated and we are left with the equation for the velocity. In problems where free surfaces are involved, we prescribe the condition on the traction at the free surface which in turn is given in the terms of the derivative of $\vec{v}$. However, when a rigid wall is involved, the condition that is enforced is the
``no-slip" condition, and prescribing the velocity seems the natural thing to do as only velocity appears in the governing equation. Also, early studies in fluid mechanics invariably appeal to semi-inverse methods where the investigators resorted to using special forms for the velocity and for the pressure; these special forms lead to ordinary differential equations for the velocity components of second or higher order (when the pressure is eliminated), which naturally calls for specification of the velocity values on the boundary. This seems wrong headed for a couple of reasons. The first, from a philosophical point of view, namely prescribing stresses in terms of kinematics as in \eqref{J1.2a} is a prescription of the ``cause" in terms of the ``effect", turning causality topsy-turvy, and second it misses the opportunity to test the aptness of the ``no-slip" boundary condition. The rest of this short paper is dedicated to discussing the second issue. 

If we were to solve the system \eqref{Jss} and \eqref{J1.4}, in general, we cannot substitute an expression for $\T$ in terms of $\nabla \vec{v}$ into \eqref{J1.4} and obtain an equation that involves $\nabla p$ and the second derivative of the velocity, we would have to solve \eqref{Jss} and \eqref{J1.4} simultaneously. Thus, it would not be natural to seek an additional condition to impermeability, that is $\vec{v}\cdot\vec{n}=0$, for the velocity. Moreover, solving \eqref{Joo} and \eqref{J1.4} entails only first derivatives in velocity.

Instead of considering \eqref{Jss} and \eqref{J1.4}, we confine ourselves to the system \eqref{J1.2b} and \eqref{J1.4} applicable to the Navier-Stokes equation and show how studying the system without assuming the ``no-slip" boundary condition allows us to evaluate its validity concerning whether the slip has to take place, thereby answering the first two questions.

It is interesting to note that the incompressible Navier-Stokes model can be expressed as a subclass of (see M\'{a}lek et al. \cite{MPrKRR})
$$   
\DD= \hat{\!\!\vec{f}}(\S), \quad \textrm{ where } \S := \T - \frac13(\operatorname{tr}\T)\I,
$$ 
an expression that meets the requirements of causality. In such a case, one would have to solve the balance of linear momentum and the constitutive relation simultaneously and not think in terms of merely specifying boundary conditions for the velocity.

\section{Illustrative examples}

We will consider five very simple problems to illustrate that the solution to these problems can be obtained without resorting to enforcing the ``no-slip" boundary condition, and moreover the methodology of obtaining the solution immediately provides a very simple and ingenious approach to determining whether there is slip!


\subsection{Poiseuille flow in a pipe} We study the flow of a Navier-Stokes fluid in a cylindrical pipe of infinite length and assume that $\vec{b}= \vec{0}$ (no gravity) and $\vec{v} = v(r) \vec{e}_z$. Then, as a consequence of the constitutive equation \eqref{J1.2b} we get $\T = -p\I + T_{rz}(\vec{e}_r \otimes \vec{e}_z + \vec{e}_z \otimes \vec{e}_r)$ where $T_{rz} = T_{rz}(r)$. The governing equations \eqref{J1.4} also imply that $p=p(z)$. 

In the \emph{classical} approach, the governing equations then reduce to only one scalar equation, namely
\begin{equation}
    p' = \mu \left( v'' + \frac{1}{r} v'\right)\,. \label{p7}
\end{equation}
Note that $p= p(z)$ and $v = v(r)$ and $p'$ means the derivative of $p$ with respect to $z$, while $v'$ and $v''$ denote the derivatives of $v$ with respect to $r$. As $p=p(z)$ and $v = v(r)$, one observes that $p' = c$, where $c$ is a constant (that is negative so that the fluid flows from from a region of higher pressure to one that is of lower pressure), and 
\begin{equation*}
    v'' + \frac{1}{r} v' = \frac{c}{\mu}\,,
\end{equation*}
which implies that 
\begin{equation}
    v(r) = \frac{cr^2}{4\mu} + c_1 \ln r + c_2\,. \qquad (c_1,c_2 \textrm{ are constants})\label{p8}
\end{equation}
The requirement that the velocity is bounded at $r=0$ implies that $c_1 = 0$. Finally, the requirement that the fluid exhibits ``no-slip" at the wall $r=R$ implies that $c_2 = - \frac{cR^2}{4\mu}$. Hence, \eqref{p8} can be expressed as
\begin{equation}
    v(r) = - \frac{c(R^2 - r^2)}{4\mu}. \label{p9}
\end{equation}

In the \emph{alternate} approach that we advocate, the structure of the quantities $\vec{v}$, $p$ and $\T$ is the same as above. This time, however, the balance of linear momentum \eqref{J1.4} is used, which gives (let us stress again that $p= p(z)$, while $v = v(r)$ and $T_{rz} = T_{rz}(r)$ and $p'$ means the derivative of $p$ with respect to $z$, while $T_{rz}'$ and $v'$ denote the derivative of $T_{rz}$ and $v$ with respect to $r$):
$$
   -p' + T_{rz}' + \frac{T_{rz}}{r} = 0 \quad \implies \quad \frac{1}{r}[r T_{rz}]' = p' \quad \implies \quad p' = c \textrm{ and } [r T_{rz}]' = r c,
$$
where $c$ is as above a negative constant (pressure gradient). Consequently,
$$
   r T_{rz} = \frac{c r^2}{2} + c_1 \quad \implies \quad T_{rz} = \frac{cr}{2} + \frac{c_1}{r}\,. 
$$ 
From the constitutive equation $T_{rz} = \mu v'$, we recover the expression \eqref{p8}, and,  again, we set $c_1 = 0$ in order to avoid the singularity at $r=0$. 

Let $Q$ be the volumetric flow rate, i.e., $Q = \int_0^R 2\pi r v(r) \, \textrm{d}r$. Using \eqref{p8} with $c_1=0$ we obtain
$$
  Q = \frac{\pi c R^4}{8\mu} + c_2 \pi R^2\,.
$$
Hence 
$$
   c_2 = \frac{1}{\pi R^2} \left[ Q - \frac{\pi c R^4}{8\mu}\right] \quad \textrm{ and } \quad v(r) =  \frac{cr^2}{4\mu} + \frac{Q}{\pi R^2} - \frac{cR^2}{8\mu}\,.
$$
Subtracting and adding the term $cR^2/(4\mu)$ we get
$$
  v(r) =  - \frac{c}{4\mu}(R^2 - r^2) + \left[\frac{Q}{\pi R^2} + \frac{c}{8\mu} R^2\right] \,.
$$
If $Q\neq - \frac{c\pi R^4}{8\mu}$, then $v(R) \neq 0$. There is slip. If $Q = - \frac{c \pi R^4}{8\mu}$, then $v(R) = 0$. There is ``no-slip". Note that the quantities $\mu$, $R$, $c$ (the pressure drop) and $Q$ can be easily measured.  

If we are certain that the pipe is a straight pipe with constant circular cross-section and sufficiently long that the end effects are negligible, then we could use the above condition to conclude whether the fluid slips or adheres at the point of contact with the solid surface. However, an obstruction in the interior of the pipe would make such a conclusion invalid. In fact, abnormality in the pressure gradient versus flow rate relationship is what a cardiologist depends upon to recognize the presence of aneurysms in blood vessels.

\subsection{Cylindrical Couette problem} Couette flow, i.e., the flow between concentric cylinders (with radii $R_{\textrm{i}}$ (inner cylinder) and $R_{\textrm{o}}$ (outer cylinder), $0<R_{\textrm{i}}<R_{\textrm{o}}$) rotating with the angular velocities $\Omega_{\textrm{i}}$ and $\Omega_{\textrm{o}}$, is characterized by the following conditions:
\begin{equation}
    \vec{v} = \omega(r) \vec{e}_{\phi} \quad \text{ and } \quad p= p(r)\,, \qquad \qquad R_{\textrm{i}} < r < R_{\textrm{o}}. \label{j.-3}
\end{equation}
We again assume that $\vec{b}= \vec{0}$.

In the \emph{classical} approach, stemming from \eqref{j.-3}, the governing equations for the Navier-Stokes fluid take the form 
$$
   p'(r) = \rho\frac{\omega^2(r)}{r} \quad \textrm{ and } \quad \frac{\mu}{\varrho} \left[ \omega'' + \frac{\omega'}{r} - \frac{\omega}{r^2} \right] = 0 
$$
with the solution given by 
\begin{align}
    \omega(r) &= Cr + \frac{D}{r}, \label{dod1}\\
    p(r) &= \varrho\left( \frac{C^2r^2}{2} - \frac{D^2}{2r^2} + 2CD \ln r\right).\label{dod2}
\end{align}
From the ``no-slip" boundary conditions for the velocity, i.e., 
\begin{equation}
    \omega(R_{\textrm{i}}) = \Omega_{\textrm{i}} \quad \textrm{ and } \quad \omega(R_{\textrm{o}}) = \Omega_{\textrm{o}},
    \label{J.bc}
\end{equation}
we conclude that 
\begin{equation}\label{dod3}
    C = \frac{\Omega_{\textrm{o}} R_{\textrm{o}} - \Omega_{\textrm{i}} R_{\textrm{i}}}{R_{\textrm{o}}^2 - R_{\textrm{i}}^2} \quad \textrm{ and } \quad D= \frac{R_{\textrm{i}}R_{\textrm{o}}(\Omega_{\textrm{i}} R_{\textrm{o}} - \Omega_{\textrm{o}} R_{\textrm{i}})}{R_{\textrm{o}}^2 - R_{\textrm{i}}^2}.
\end{equation}
Then 
\begin{equation*}
    \omega(r) = \frac{\Omega_{\textrm{o}} R_{\textrm{o}} - \Omega_{\textrm{i}} R_{\textrm{i}}}{R_{\textrm{o}}^2 - R_{\textrm{i}}^2}r + \frac{R_{\textrm{i}}R_{\textrm{o}}(\Omega_{\textrm{i}} R_{\textrm{o}} - \Omega_{\textrm{o}} R_{\textrm{i}})}{R_{\textrm{o}}^2 - R_{\textrm{i}}^2} \frac{1}{r},
\end{equation*}
and we get corresponding formula for $p$ from \eqref{dod2} and \eqref{dod3}. 


In the \emph{alternate} approach, we start with governing balance equations that take the form 
\begin{equation}
    \varrho\frac{\omega^2(r)}{r} = p' \quad \textrm{ and } \quad \frac{(r^2 T_{r\phi})'}{r^2} = \frac{2}{r} T_{r\phi} + T_{r\phi}' = 0 \label{Jo.-1}
\end{equation}
together with the constitutive equation of the form
\begin{equation}
     T_{r\phi} = \mu  \left[ \omega' - \frac{\omega}{r}\right]. \label{Jo.-2}
\end{equation}
{Let us assume that we know the applied torques at the inner and outer cylinders, i.e.,
\begin{equation}
    T_{r\phi}(R_{\textrm{i}}) = M_{\textrm{i}} \quad \textrm{ and } \quad T_{r\phi}(R_{\textrm{o}}) = M_{\textrm{o}}. \label{Jo.-4}
\end{equation}
Then it follows from the second equation 
in \eqref{Jo.-1} and these boundary conditions that 
$$
   T_{r\phi}(r) = \frac{M_{\textrm{o}} R_{\textrm{o}}^2}{r^2} = \frac{M_{\textrm{i}} R^2_{\textrm{i}}}{r^2}.
$$
It means that in order to get the solution of the form \eqref{j.-3} the data $R_{\textrm{i}}$, $R_{\textrm{o}}$, $M_{\textrm{i}}$ and $M_{\textrm{o}}$ have to satisfy the following compatibility condition 
\begin{equation}\label{dod4}
  M_{\textrm{o}} R_{\textrm{o}}^2 = M_{\textrm{i}} R_{\textrm{i}}^2.
\end{equation}
Next, using  \eqref{Jo.-2} we observe from
$$
\mu \left(\frac{\omega}{r}\right)' = \frac{1}{r} \mu \left[\omega' - \frac{\omega}{r}\right] = \frac{M_{\textrm{o}} R_{\textrm{o}}^2}{r^3}
$$
that 
\begin{equation}
    \omega(r) = \frac{1}{\mu} \left[\beta r - \frac{M_{\textrm{o}}R_{\textrm{o}}^2}{2r}\right]\,.\label{Jo.-5}
\end{equation}}
To fix the coefficient $\beta$, we use the equation for $p$ and assume that 
\begin{equation}
    p(R_{\textrm{i}})=p_{*}, \label{J.-5}
\end{equation}
where $p_*$ is a pressure that could be measured by a pressure transducer at the inner part of the inner cylinder. With $C=\frac{\beta}{\mu}$ and $D= - \frac{M_{\textrm{i}}R_{\textrm{i}}}{2}$ (consequence of \eqref{Jo.-5} and \eqref{dod4}), the general formula for the pressure leads to the following quadratic equation for $\beta$, namely 
$$
  \frac{R_{\textrm{i}}^2}{2\mu^2} \beta^2 - \frac{M_{\textrm{i}} R_{\textrm{i}}^2 \ln R_{\textrm{i}}}{\mu^2} \beta  - \frac{M_{\textrm{i}^2}R_{\textrm{i}}^2}{8\mu^2} = \frac{p_*}{\varrho} \quad \implies \quad R_{\textrm{i}}^2 \beta^2 - 2 M_{\textrm{i}} R_{\textrm{i}}^2 (\ln R_{\textrm{i}}) \beta  - \frac{M_{\textrm{i}}^2R_{\textrm{i}}^2}{4} = \frac{2\mu^2 p_*}{\varrho}.
$$ 
We do not provide the explicit formula for $\beta$, but note that under certain conditions on the data, the solution of the above equation can have two or one or no solution. Assuming we fix $\beta$, referring again to \eqref{Jo.-5}, we observe that if $2\beta = M_{\textrm{i}}$ then there is ``no-slip" on the inner cylinder (otherwise the fluid slips there), while if $2\beta = M_{\textrm{o}}$ then there is ``no-slip" on the outer cylinder. Note that under certain conditions we have two values of pressures that can adjust both these conditions.  

Alternatively, we could fix $\beta$ by prescribing the volumetric flow rate $Q$ across a cross section. We show the calculation with the caveat that for the cylindrical Couette flow $Q$ is not easy to measure. 
Using \eqref{Jo.-5}, it follows from 
$$
Q= \int_{R_{\textrm{i}}}^{R_{\textrm{o}}} \omega(r) \, \textrm{d}x $$
that 
$$
\omega(r) = \frac{1}{\mu} \left[\frac{2\mu Q + M_{\textrm{o}}R_{\textrm{o}}^2 \ln \frac{R_{\textrm{o}}}{R_{\textrm{i}}}}{R_{\textrm{o}}^2 - R_{\textrm{i}}^2} r - \frac{M_{\textrm{o}}R_{\textrm{o}}^2}{2r}\right]\,.
$$
Now, if $M_{\textrm{o}} = 2\frac{2\mu Q + M_{\textrm{o}}R_{\textrm{o}}^2 \ln \frac{R_{\textrm{o}}}{R_{\textrm{i}}}}{R_{\textrm{o}}^2 - R_{\textrm{i}}^2}$, then $\omega(R_{\textrm{o}})= 0$, i.e., there is ``no-slip". If the condition is not fulfilled there is slip. Similarly, using also the compatibility condition \eqref{dod4}, if 
$M_{\textrm{i}} = 2 \frac{2\mu Q + M_{\textrm{i}}R_{\textrm{i}}^2 \ln \frac{R_{\textrm{o}}}{R_{\textrm{i}}}}{R_{\textrm{o}}^2 - R_{\textrm{i}}^2}$, then $\omega(R_{\textrm{i}})=0$ and the fluid adheres to the inner cylinder. Otherwise, there is slip.

\subsection{Plane Poiseuille flow} We consider the flow that takes place between two parallel plates located at $y=0$ and $y=h$ and assume that $\vec{b}= \vec{0}$ (no gravity). Furthermore, we assume that $\vec{v} = u(y)\vec{i}$. 

In the \emph{classical} approach, the second and third equation of \eqref{J1.6} imply that $p=p(x)$, while the first equation reduces to 
$$
   -\frac{\textrm{d}p}{\textrm{d}x} + \mu \frac{\textrm{d}^2u}{\textrm{d}^2 y} = 0\,
$$ 
which implies that 
$$
  p(x) = cx + b \quad\textrm{ and } \quad u(y) = \frac{c}{2\mu} y^2 + dy + e\,,  
$$
where $b$, $c$, $d$ and $e$ are constants, $c$ being negative so that the fluid flows from a region of higher pressure to one that is of lower pressure. Requiring ``no-slip" boundary conditions on the plates, i.e., $u(0) = u(h) = 0$ one concludes that 
\begin{equation}
      u(y) = \frac{C}{2\mu} y(y-h)\,.\label{p5} 
\end{equation}
 
In the \emph{alternate} (new) approach, we first observe that the assumption $\vec{v} = u(y)\vec{i}$ and the constitutive equation \eqref{J1.2b} imply that 
\begin{equation}
\T = \begin{pmatrix} -p(x,y,z) & \tau(y) & 0 \\
\tau(y) & -p(x,y,z) & 0 \\ 0 & 0 & -p(x,y,z) \end{pmatrix}\,.\label{dod5}
\end{equation}
Then the second and third equation in \eqref{J1.6} lead to $p = p(x)$, while the first equation of \eqref{J1.6} then gives 
$$
   -\frac{\textrm{d}p}{\textrm{d}x} + \frac{\textrm{d}\tau}{\textrm{d}y} = 0. 
$$
Consequently, 
$$
  p(x) = cx + b \quad \textrm{ and } \quad \tau(y) = cy + d\,,  
$$
where again $b$, $c$ and $d$ are arbitrary constants, $c$ being negative. The constitutive equation $\tau = \mu u'$ then leads to 
\begin{equation}
   u(y) = \frac{c}{2\mu} y^2 + \frac{d}{\mu} y + e\,.\label{p2}
\end{equation}
Now, we impose two new conditions, namely 
\begin{equation}
   u(h) = u(0) \quad (\textrm{symmetric velocity profile}) \qquad \textrm{ and } \quad \int_0^h u(y) \, \textrm{d}y = Q.\label{p3} 
\end{equation}

Applying the first condition from \eqref{p3} on the formula given in \eqref{p2}, we observe that 
$$
  \frac{c}{2\mu} h^2 + \frac{d}{\mu} h + e = e \quad \implies \quad d = - \frac{ch}{2}.
$$
Substituting this in \eqref{p2} and using the second condition in \eqref{p3} we obtain
$$
  Q = -\frac{c}{12\mu} h^3 + eh \quad \implies \quad e = \frac{Q}{h} + \frac{ch^2}{12\mu}\,.
$$
Hence 
$$
   u(y) = \frac{c}{2\mu} y^2 - \frac{ch}{2\mu} y + \frac{ch^2}{12\mu} + \frac{Q}{h}\,.
$$
This implies that 
$$
  u(0) = u(h) =  \frac{ch^2}{12\mu} + \frac{Q}{h}\,.
$$
We conclude that if $Q\neq -\frac{ch^3}{12\mu}$, then $u(0) \neq 0$ and $u(h) \neq 0$ as well. That means that the fluid is slipping. If, however,  $Q = -\frac{ch^3}{12\mu}$, then $u(0)=0$, $u(h)=0$ and the solution takes the form \eqref{p5}, as obtained above using the classical approach. 

\begin{figure}[h]
\tikzmath{\l = 8; \b =4; \h =0.4;}
\includegraphics[width=0.6\linewidth]{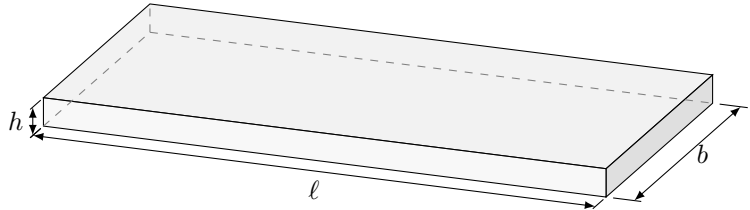}
\caption{Cylinder of rectangular crossection with dimensions $\ell, b, h$}
\label{f1}
\end{figure}
As an experiment between two parallel plates is not really feasible, the above calculation has been provided to show that one can obtain the solution to the flow problem without having to specify the ``no-slip" boundary condition. If the fluid is flowing through a cylinder of rectangular cross-section due to a pressure gradient (see Fig. \ref{f1}) and if $b\gg h$, $\ell\gg h$, so that end effects can be neglected, we can assume, except regions very close to the vertical walls, the fluid is undergoing plane Poiseuille flow as a reasonable approximation and such an experiment can be performed and $Q$ can be measured. The point to be made is that such an approach of not specifying the "no-slip" condition, but using other alternative conditions could be used in other problems wherein one could indeed carry out experiments, and conclude the validity of the "no-slip" condition.

\subsection{Plane Couette flow.} The flow takes place between two parallel plates located at $y=0$ and $y=h$. We shall assume that $\vec{b}= \vec{0}$ (no gravity), $\vec{v} = u(y)\vec{i}$ and $p$ is constant. It then follows from \eqref{dod5} that $\diver \T = (\tau', 0, 0)$.


We first recall the \emph{classical} approach. This is characterized by the fact that the upper plate moves with the constant velocity $V$. Under these circumstances, the problem \eqref{J1.5}-\eqref{J1.6} reduces to $u''(y) = 0$, which implies that $u(y) = Cy+D$. If $u(0)= 0$ and $u(h)=V$, then $u(y) = \frac{V}{h}y$. 

In the \emph{alternate} (new) approach we assume that a shear stress $\tau_{\textrm{app}}$ is applied on the fluid by the moving plate in contact with the fluid (this being equal and opposite to the shear stress exerted by the fluid on the plate), i.e.,
\begin{equation}
    \tau(h) = \tau_{\textrm{app}}. \label{p1}
\end{equation}
The governing system of equations \eqref{J1.2b} and \eqref{J1.4} reduces to
$$
\tau = \mu u' \quad \textrm{ and } \quad  \tau' = 0\,.
$$ 
This together with the boundary condition \eqref{p1} yields
$$
 \tau(y) = \tau_{\textrm{app}} \quad \textrm{ and } \quad u(y) = \frac{\tau_{\textrm{app}}}{\mu} y + D\,,
$$
where $D$ is a constant.

At this juncture, let us assume that we can measure the volumetric flow rate $Q$ and assume that it is given. Thus, 
$$
  Q= \int_0^h u(y) \, \textrm{d}y = \frac{\tau_{\textrm{app}}}{2\mu} h^2 + D h\,,
$$
which leads to 
$$
  D = \frac{1}{h} \left[Q - \frac{\tau_{\textrm{app}}}{2\mu} h^2\right].
$$
Thus 
$$
  u (y) = \frac{\tau_{\textrm{app}}}{\mu} y + \frac{1}{h} \left[Q - \frac{\tau_{\textrm{app}}}{2\mu} h^2\right].
$$ 
Notice that $u(0) = \frac{1}{h} \left[Q - \frac{\tau_{\textrm{app}}}{2\mu} h^2\right]$. If $\left[Q - \frac{\tau_{\textrm{app}}}{2\mu} h^2\right]\neq 0$, then we can conclude that there has to be slip! If $\left[Q - \frac{\tau_{\textrm{app}}}{2\mu} h^2\right] =  0$, then there is ``no-slip" at the lower plate. 

One can do the same at the upper plate (located at $y=h$). As  
$$
  u(h) = \frac{\tau_{\textrm{app}}}{\mu} h + \frac{1}{h} \left[Q - \frac{\tau_{\textrm{app}}}{2\mu} h^2\right] = \frac{\tau_{\textrm{app}}}{2\mu} h + \frac{Q}{h},
$$
we observe that if the speed $V$ of the upper plate associated with the applied shear stress $\tau_{\textrm{app}}$ is such that 
$$
  \frac{\tau_{\textrm{app}}}{2\mu} h + \frac{Q}{h} \neq V\,,
$$
then we can conclude that there is slip! 

As in the previous case, the main purpose of the calculation is to show that one does not need the "no-slip" boundary condition to obtain a solution to the flow problem. 

\subsection{Flow down an inclined plane due to gravity} In the co-ordinate system associated with the inclined plane, the gravitational force takes the form $\varrho\vec{b} = (\varrho g \sin\theta, -\varrho g \cos\theta, 0)$, where $g$ is the acceleration due to gravity and $\theta$ is the angle of inclination.

In the \emph{classical} approach, starting from the assumption $\vec{v}=u(y)\vec{i}$ and $p=p(y)$ one deduces from the governing equations \eqref{J1.2b}-\eqref{J1.5} that the only non-diagonal element of $\T$
is $T_{xy} = T_{yx} = \tau(y)$.  The equations stemming from \eqref{J1.6} takes the form 
\begin{align}
    \mu u'' + \varrho g \sin \theta &= 0\,, \label{1c} \\
    -p' - \varrho g \cos\theta &= 0\,. \label{1d}
\end{align}
After integrating \eqref{1c} we obtain ($\ell$ is a constant)
\begin{equation}
    \mu u' = -\varrho g (\sin\theta) y + \mu \ell\,.\label{1e}
\end{equation}
This implies that ($m$ is yet another constant)
\begin{equation}
    u(y) = -\frac{\varrho g (\sin\theta)}{2\mu} y^2 + \ell y + m\,. \label{1f}
\end{equation}
At the free surface, one assumes that $\T\vec{n}$ vanishes. Using \eqref{J1.2b} and the above assumptions $\T\vec{n} = \T(0,1,0) = (\tau(h), -p(h), 0) = (\mu u'(h), -p(h), 0)$. Consequently, $\T\vec{n}=\vec{0}$ implies that 
$$
u'(h) = 0\quad \textrm{ and } \quad p(h)=0\,.
$$
The first condition together with \eqref{1e} leads to $\ell= \frac{\varrho g (\sin\theta) h}{\mu}$. Hence,
\eqref{1f} takes the form 
$$
  u(y) = \varrho g \sin\theta ({h}/{2} -y) y + m\,.
$$
The required ``no-slip" condition on the bottom, i.e., $u(0)=0$ gives $m=0$ and thus 
$$
  u(y) = \varrho g \sin\theta ({h}/{2} -y) y\,.
$$
Note that the condition $p(h)=0$ together with \eqref{1d} implies 
\begin{equation}
  p(y) = \varrho g \cos \theta (h-y)\,. \label{1.6}
\end{equation}

In the \emph{alternate} (new) approach, we have $\vec{v} = u(y) \vec{i}$, $p=p(y)$ and $\tau = \tau(y)$.
The balance of linear momentum \eqref{J1.4} implies 
\begin{align}
    \tau' + \varrho g \sin \theta &= 0\,, \label{1cc} \\
    -p' - \varrho g \cos\theta &= 0\,. \label{1dd}
\end{align}
After integrating \eqref{1cc} we obtain ($\tilde\ell$ is a constant)
\begin{equation}
    \tau(y) = -\varrho g (\sin\theta) y + \mu \tilde\ell\,.\label{1ee}
\end{equation}
The constitutive equation \eqref{J1.2b} implies that ($\tilde m$ is a constant)
\begin{equation}
    u(y) = -\frac{\varrho g (\sin\theta)}{2\mu} y^2 + \tilde\ell y + \tilde m\,. \label{1ff}
\end{equation}
At the free surface, we assume that $\T\vec{n} = (\tau(h), -p(h), 0)$ vanishes, i.e., 
$$
  \tau(h) = 0\quad \textrm{ and } \quad p(h)=0\,.
$$
The first condition together with \eqref{1ee} gives $\tilde\ell= \frac{\varrho g (\sin\theta) h}{\mu}$. Hence,
\eqref{1ff} takes the form 
$$
  u(y) = \frac{\varrho g \sin\theta}{\mu} ({h}/{2} -y) y + \tilde m\,.
$$
Assuming that we know the volumetric flow rate $Q$, we can fix $\tilde m$. As $Q= \int_0^h u(y)\, \textrm{d}y$, we conclude from the above formula for $u$ that 
$$
  Q = \int_0^h \left[ \frac{\varrho g \sin\theta}{\mu} ({h}/{2} -y) y + \tilde m \right]\, \textrm{d}y = \frac{\varrho g (\sin\theta) h^3}{3 \mu} + \tilde m h\,,
$$
which implies that 
$$
\tilde m = \frac{Q}{h} - \frac{\varrho g (\sin\theta) h^2}{3 \mu}\,. 
$$
Clearly, if $Q\neq \frac{\varrho g (\sin\theta) h^3}{3 \mu}$, then $\tilde m \neq 0$ and consequently $u(0) \neq 0$! The fluid has the slip along the inclined plane. On the other hand, if $Q = \frac{\varrho g (\sin\theta) h^3}{3 \mu}$, then $\tilde m = 0$ and $u(0)=0$. 

\section{Conclusion}


In this paper, we have articulated a methodology for obtaining solutions to flow problems of the Navier-Stokes fluid without appealing to the ``no-slip" boundary conditions and we have also advanced a procedure for testing the validity of the ``no-slip" boundary condition based on easily experimentally measurable quantities. On knowing that the fluid slips at the solid boundary, we can make additional assumptions for the nature of the slip, say Navier's slip etc., and in fact determine the extent of slip. We intend to address this in a forthcoming study. 

In general, we show that it might be best, in the absence of certain knowledge concerning the boundary condition for the velocity at a solid boundary to use the system of equations consisting of the balance laws as well as the constitutive relation simultaneously as in the case of implicit constitutive relations.

\providecommand{\bysame}{\leavevmode\hbox to3em{\hrulefill}\thinspace}
\providecommand{\MR}{\relax\ifhmode\unskip\space\fi MR }
\providecommand{\MRhref}[2]{%
  \href{http://www.ams.org/mathscinet-getitem?mr=#1}{#2}
}
\providecommand{\href}[2]{#2}

\end{document}